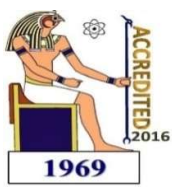

**Delta Journal of Science**

Available online at https://djs.journals.ekb.eg/

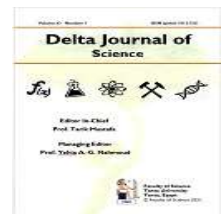

Research Article                                                                                                           **PHYSICS**

# Kinetic Simulation of He radio frequency capacitive coupled plasma

M. Shihab[1,2,*], A. Elbadawy[1], M. S. Afify[3], and N. El-Siragy[1]

[1] Tanta University, Faculty of Science, Physics Department, Tanta 31527, Egypt.
[2] Academy of Scientific Research and Technology (ASRT), Cairo, Egypt.
[3] Department of Physics, Faculty of Science, Benha University, Benha, P.O. Box 13518, Egypt.
[*]Mohammed.shihab@science.tanta.edu.eg

| KEY WORDS | ABSTRACT |
|---|---|
| Radio frequency sheaths, He discharge, Ion energy and angular distribution, Electron energy distribution, Power dissipation. | Radiofrequency capacitively coupled plasma is studied theoretically using a Particle-in-Cell code. For He discharge, the time-averaged sheaths are in the range of few centimeters. The sheath potential, ion, and electron energy and angular distributions, discharge current, and dissipated power depend on the driven potentials and frequencies. Increasing the amplitude of the high radio frequencies increases the bulk density and the sheath potential and, consequently, increases the plasma processing rate. Increasing the intermediate radio frequency amplitude allows a wider sheath with a broad ion energy distribution and a narrower ion angular distribution. Changing the amplitude and the phase shift between driven frequencies provide different energies and angular distribution allowing performing various processes. The interplay between the sheath and bulk dynamics in the intermediate radiofrequency regime and the high-frequency regime may excite harmonics in the discharge current. |

## 1. Introduction

Low-temperature plasma has a great potential for numerous applications in the growth and processing of nanomaterials and the fabrication of microelectronics, e.g., carbon nanotubes, nanowires, thin-film depositions, and anisotropic etching of metallic, semiconductor, and dielectric materials. The energy of incident ions on substrates determines the process type and the flux of ions determines the rate of the process. In this contribution, we study Helium (He) discharge utilizing the Particle-In-Cell technique. He is an inert gas. Its chemistry is simple and could be used to host different gas compositions - such as $O_2$, $N_2$, $CF_4$, $CH_4$, $H_2O$- without affecting their chemistry. Here, we try to reveal the effect of the amplitude of driven radio frequencies and their phase shift on the discharge dynamics, the ion energies and the ion angular distribution at electrodes, the electron distribution, and the dissipated power in the plasma. The driven frequencies are 60 MHz and 1 MHz. Tailoring the driven potential or driven electrodes with different radio frequencies is one of the hot topics of research nowadays [7, 8, 9].

In the next section, we give a short overview of the Particle-in-Cell technique, then in section 3 we introduce our results and close the manuscript with a conclusion in section 4.

## 2. Particle-In -Cell

In the approach of modeling particles in a cell, Particle-in-Cell (PIC) has been widely utilized as a computational method to understand and forecast the behavior of plasma [1, 2]. An interpolation approach is used to collect charge and current density on a spatial mesh. The simulation domain is discretized into $k_{th}$ grids as shown in figure (1). On grids, the field equations are solved. Interpolation is used to determine the force exerted on the super-particles. Each super-particle represents $10^3$ t0 $10^6$ of real particles. In Fig. (2), the PIC simulation's computational cycle is depicted. The indices i and k are used to designate quantities evaluated for particles and grid points, respectively. There are four steps in the computing cycle (excluding the Monte-Carlo collision).

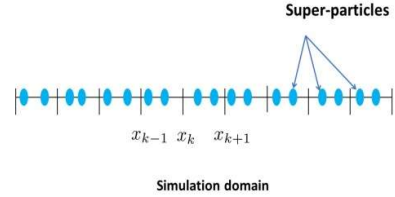

**Fig. (1):** The discretization of the simulation domain into $k_{th}$ grids.

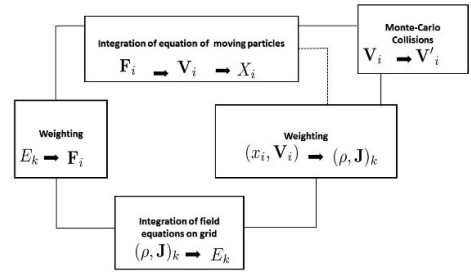

**Fig (2):** A schematic flow chart of PIC modules. The dashed lines are to short cut Monte-Carlo calculations for collisionless plasma.

The particles' locations are computed initially. The force effects on and the acceleration of each particle are computed, then the particles' locations and velocities are updated in the first step. The charge density is then calculated at the grid points using the charge of each particle and the current in the second step, which is known as weighting. The third stage integrates the Poisson's equation to obtain the electric potential and uses a finite difference technique to compute the electric field E on the grid. In the final stage, the electric field at surrounding grid points is used to calculate the force on each particle. Because PIC algorithms can use Monte-Carlo methods to simulate collisions between charged particles and neutral atoms, an extra stage in the computing cycle is added, see, Fig. (2). Considering

Kinetic Simulation of He radio frequency capacitively coupled plasma                                                                    91collision between particles is time consuming. The null collision method is employed to perform simulation in a proper time. This method involves selecting a particle and comparing the relative chance of a collision to a random number to see if one really occurs. In low temperature plasma, the degree of ionization is small, hence, only electron-neutral and ion-neutral models were used for these simulations. Scalability is crucial because it connects simulations to the real world of plasma. The values supplied to parameters such as grid spacing, time step, and super-particle density are critical since they determine the simulation's speed and accuracy. A few aspects must be considered before performing a PIC simulation of the plasma in order to avoid unphysical outcomes.

In the simulation, the number of super-particles $N_p$ should be substantially more than the number of grid cells $n_g$, i.e. $N_p \gg n_g$. This is to ensure that, on average, each grid cell includes multiple particles during the simulation. The simulation will be noisy if the number of particles is too low. The grid cell size $\Delta x$ should be on the order of the Debye length. The Debye length is the longest distance over which individual particle Coulomb forces are significant. If the grid spacing is reduced, we will be unable to eliminate short-range particle interactions, which are irrelevant to the plasma's overall behavior. Important electric field gradients can be neglected if the spacing is made bigger than Debye length, because the fields are only determined at the grid points. Furthermore, because grid cells affect particle size, larger grid cells will produce unphysical results. In addition, the time step $\Delta t$ must be less than the period of plasma oscillations and $\Delta t$ should be small enough to allow stable and precise integration of the particle equations of motion, as well as correct reproduction of particle oscillations.

Particles passing across cell borders generate density fluctuations, which are then transmitted to the potentials and electric fields. If the fluctuations are lower than the magnitude of the applied potentials and do not cause unstable behavior, they can be ignored. Electrons are extremely sensitive to field fluctuations, which can lead to an unphysical increase in electron energy. Ions are slower to respond to fields; therefore, transitory fluctuations have no effect on them. If the artificial rise in electron energy grows great enough, it might produce excess ionization, which raises plasma density, which magnifies the heating, resulting in an exponential increase in plasma density, and the simulation eventually breaks down. Another issue is the loss of resolution in low-density areas, such as sheaths. As a result, the number of particles in a super-particle must be reduced, i.e. smaller super-particles must be used. For more details about PIC simulation, please read [10, 11, 12, 13].

## 3. Results and discussion

Here, we study employing Particle-in-Cell (PIC) ions' and electrons distributions in radio frequency capacitively coupled plasmas (RF-CCPs), where electrodes are biased with two radio frequencies, 1 MHz and 60 MHz. The code is benchmarked, so its predictions are trustful [14]. For simulation, a geometrically symmetric reactor is chosen. The distance between electrodes is 15 cm. The time and space are discretized in a way to avoid numerical instabilities. The simulation is repeated for 500 RF cycles of the 60 MHz. The time step is 1/600 from the periodic time of the 60 MHz. The distance between the two electrodes are discretized into 259 grids. The 1 MHz is comparable or smaller than typical ion plasma frequency in RF-CCPs, while the 60 MHz is much higher than the ion plasma frequency. The 1 MHz allow ions to respond partially to the instantaneous electric fields. On contrary, the 60 MHz compel ions to respond to the time averaged field. The driven potentials $V = V_1 \sin(2\pi\, 60\text{MHz}\, t) + V_2 \sin(2\pi\, \text{MHz}\, t + \theta)$. Based on the amplitude of the driven potentials, three cases are considered: In case (1), $V_1 = V_2 = 250V$, where the effect of both frequencies is supposed to be equal. In case (2), $V_1 = 100V$ and $V_2 = 400V$; the plasma is mainly driven by the intermediate frequency. In case (3), $V_1 = 400V$ and $V_2 = 100V$; the plasma is ignited by the high frequency. The simulations are carried out twice for the three cases when the phase shift ($\theta$) is zero, corresponding results are displayed as solid lines. When the phase shift $\theta$ is $\pi/2$, results are presented via dashed lines.

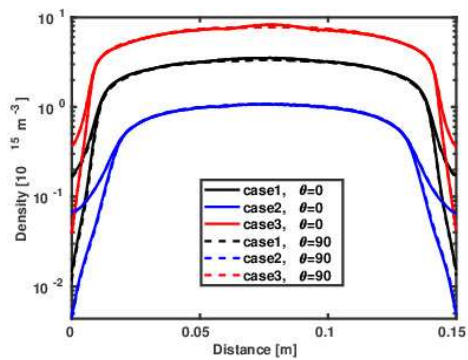

Fig. 3. The plasma density between the two electrodes.

Let first discuss the results when there is no phase shift. Close to electrodes, quasineutrality breaks down and RF sheaths are formed due to the escape of electrons from the discharge volume into electrodes.

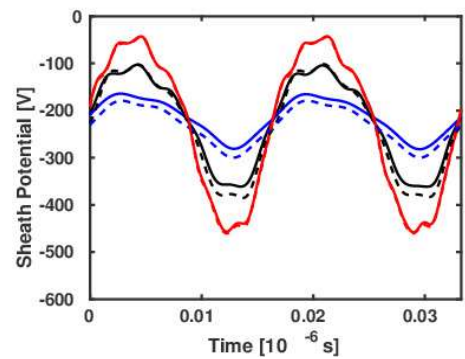

Fig. 4. The sheath potential for different plasma simulation cases shown in Fig. 3.

The time averaged density of the left sheath for three cases are shown in Figure 3. Case (1), case (2), and case (3) are represented with black, blue, and red solid lines, respectively. For each case, the upper line presents the ion density and the lower gives the electron density. The minimum sheath width is belonging to case (3), where, the amplitude of the high frequency signal is larger than the intermediate frequency. The power of the high frequency signal is mainly dissipated in the plasma bulk, therefore, the bulk density increases and providing a larger ion flux to the plasma sheath. When the amplitude of the high frequency is larger than the amplitude of intermediate frequency, i.e., case 2, the time averaged sheath is in the order of 1 cm. For case 3, the time averaged sheath width is roughly 2 cm. These large sheaths suggested that typical Ar capacitively coupled plasma reactors with a gap size of 5 cm or less may be not suitable for He discharges. For plasma etching, deposition, and sputtering at low pressures, the mean free path is large, and the thickness of the bulk may be not enough to ignite the plasma via electron-neutral background collisions.

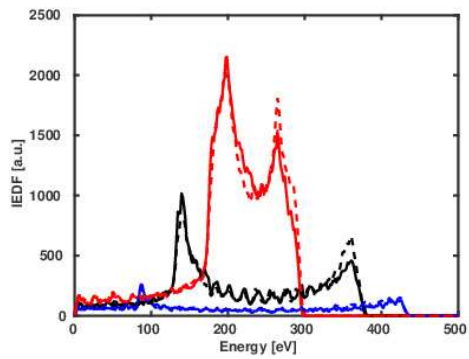

**Fig. 5.** The ion energy distribution function at the left electrode for different plasma simulation cases shown in Fig. 3.

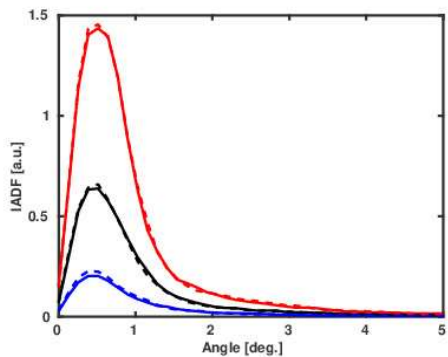

**Fig. 6.** The ion angular distribution function at the left electrode for different plasma simulation cases shown in Fig. 3.

The corresponding sheath potentials are present in Figure 4 with the same legend as in Fig. 3. If we look only to solid lines, case (1) which is dominated by the potential of the high frequency has the largest peak to peak sheath potential. The ion energy distributions (IED) are shown in Fig. 5. Case (3) has the narrowest distribution shown as red solid line. The broadest IED is obtained when the intermediate frequency potential is dominant as in case (2). The corresponding ion angular distribution function (IADF) is shown in Figure 6. The highest peak of the IAD belongs to case (3) and the lowest one is due to case (2). From previous calculations, increasing the ion flux, the sheath potential, and decreasing the sheath width allow high peak of the ion angular distribution [8]. This matched very well with case (3), please investigate Figure 1 and Fig. 2. Considering a phase shift (θ) of π/2, a slight increase in the sheath width of case (1) is observed. However, the two sheaths for case (2) and case (3) still roughly the same. The bulk densities are not sensitive to the phase shift for all cases. The sheath potential for case (3) is the same with and without a phase shift. The red and dashed solid lines are almost identical. Also, for case (1) and case (2), the sheath potentials with and without the phase shift are comparable. Therefore, the IAD and IED for all cases are almost identical.

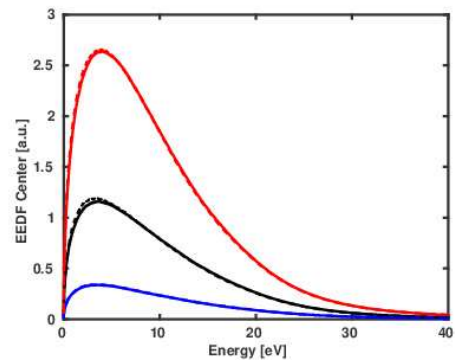

**Fig.7.** The electron energy distribution function at the center of the dischargefor different plasma simulation cases shown in Fig. 3.

Also, as shown in Fig. 7 and 8, the electron energy distribution affected by changing the driven potentials. Increasing the amplitude of the high frequency component increases the height and the width of the electron energy distribution. On contrary, intermediate frequencies allow narrower electron energy distribution. At lower radio frequencies electrons may be able to enter nano structures etched in substrates to neutralize positive charges on the etched trench surfaces [7]. The electron energy distribution at the center of the discharge is not affected by the phase shift. Only for cases (1) and (2) at the electrode, the height and the width of the distribution increased by adding a phase shift of π/2.

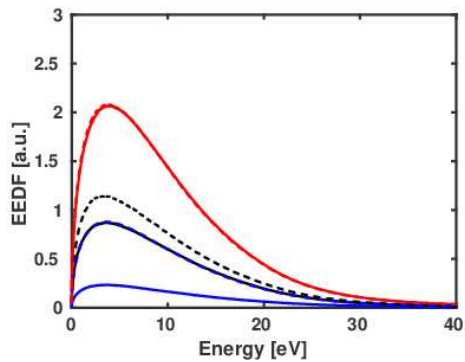

**Fig. 8.** The electron energy distribution function at the left electrode for different plasma simulation cases shown in Fig. 3.

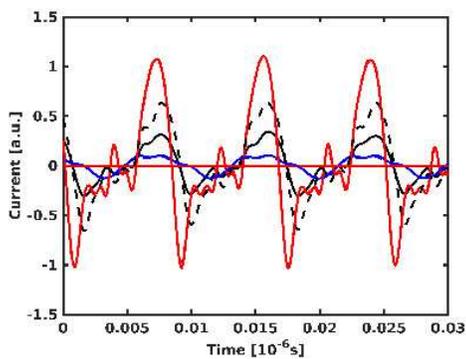

**Fig. 9.** The current passing through the discharge for different plasma simulation cases shown in Fig. 3.

Also, to reveal possible resonances between the plasma bulk and the sheath, the current is shown in Fig. 9. For case (2) and case (3), the phase shift has no effect on the passing current. But for case (1), the amplitude of the current increases by increasing the phase shift. When there is not nonlinear interaction between the sheath and bulk dynamics, the Fourier analysis of the current should only display a two component in the frequency domain; i.e., 1 MHz and 60 MHz. As could be seen in Fig. 8, other component is generated in the plasma. The amplitude of these components are a function of the driven potentials and phase shifts.

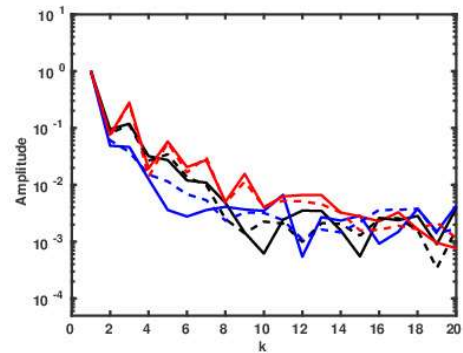

**Fig. 10.** The Fourier component of the current passing through the discharge for different plasma simulation cases shown in Fig. 3.

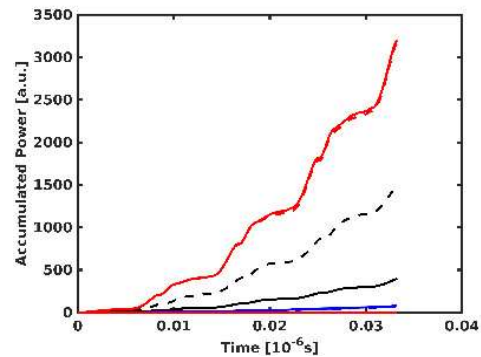

**Fig. 11.** Accumulated power as a function of time for different plasma simulation cases shown in Fig. 3.

The accumulated power is depicted in Fig. 11. The accumulated power increases by increasing the amplitude of the high frequency. It is not sensitive to the phase shift when the discharge is dominantly driven by high or intermediate frequency. However, when both amplitudes are equal, the phase shift has an effect. Plasma series resonance is responsible for the generation of new harmonics which affect the dissipated power [15, 16].

## 4. Summary

The discharge dynamics have been found to be controlled via tailoring the driven potential. For He discharge, the time-averaged sheaths are large, and typical Ar discharge RF-CCP reactors maybe not appropriate for He discharge. Increasing the amplitude of the high radio frequencies has been found to increase the bulk density and the sheath potential. Also, it has been found to accelerate ions in the plasma sheath with energies around the time-averaged values. On contrary, increasing the amplitude of the intermediate radiofrequency has been found to provide a wider sheath with a wider ion energy distribution and a narrower ion angular distribution. The height and the width of the electron energy distribution function was a function of the amplitude of the driven potentials. The plasma series resonance has been found to generate new harmonics in the discharge current and enhances the dissipated power to generate the plasma.

## 5. Acknowledgment

The authors thank the great discussion with T. Mussenbrock (Ruhr-University Bochum) and his YAPIC code is acknowledged. This project was supported financially by the Academy of Scientific Research and Technology (ASRT), Egypt, Grant No 6742. ASRT is the 2nd affiliation of this research.

# محاكاة حركية لبلازما الهليوم فى مفاعلات البلازما السعوية فى مدى موجات الراديو

محمد شهاب[1,2]، آية البدوى[1]، محمود سعد عفيفى[3]، نبيل السراجى[1]

[1]جامعة طنطا، كلية العلوم، قسم الفيزياء، طنطا، مصر.
[2]أكاديمية البحث العلمى والتكنولوجيا، القاهرة، مصر.
[3]قسم الفيزياء، كلية العلوم، جامعة بنها، بنها، مصر.

**تم دراسة المفاعلات السعوية فى مدى ترددات الراديو لبلازما الهليوم باستخدام نموذج جسم فى خلية. وأوضحت الدراسة أن خصائص البلازما المتكونة تتوقف على الجهود الكهربية المطبقة والترددات المستخدمة وكذلك فرق الطور بين هذه الترددات. يمكن من خلال تغيير جهود الترددات وقيمها وفرق الطور بينها التحكم فى طاقات الايونات والالكترونات الساقطة على الاقطاب والشرائح ومن ثم التحكم فى معدلات و نوع المعالجة بواسطة البلازما.**